\documentclass[usenatbib]{mnras}
\usepackage{microtype}
\usepackage{newtxtext,newtxmath}

\usepackage[T1]{fontenc}
\usepackage{ae,aecompl}
\usepackage{color}

\usepackage{graphicx}	
\usepackage{amsmath}	
\usepackage{amssymb}	
\definecolor{violet(ryb)}{rgb}{0.53, 0.0, 0.69}
\definecolor{scarlet}{rgb}{1.0, 0.13, 0.0}

\newcommand{\bdw}[1]{\textbf{\textcolor{scarlet}{BDW: #1}}}

\graphicspath{ {figures/} }

\title{ {FSD}: Frequency Space Differential measurement of CMB spectral distortions}
\author[Mukherjee, Silk \& Wandelt]{
Suvodip Mukherjee,$^{1,2,3}$\thanks{E-mail: smukherjee@flatironinstitute.org}
Joseph Silk,$^{2,3,4,5}$\thanks{E-mail: joseph.silk@physics.ox.ac.uk}
Benjamin D. Wandelt$^{1,2,3,6}$\thanks{E-mail: bwandelt@iap.fr}
\\
$^{1}$Center for Computational Astrophysics, Flatiron Institute, 162 5th Avenue, 10010, New York, NY, USA\\
$^{2}$Institut d'Astrophysique de Paris\\ 98bis Boulevard Arago, 75014 Paris, France\\
$^{3}$Sorbonne Universités, Institut Lagrange de Paris\\ 98 bis Boulevard Arago, 75014 Paris, France\\
$^{4}$The Johns Hopkins University, Department of Physics \& Astronomy, \\ Bloomberg Center for Physics and Astronomy, Room 366, 3400 N. Charles Street, Baltimore, MD 21218, USA\\
$^{5}$Beecroft Institute for Cosmology and Particle Astrophysics, University of Oxford, Keble Road, Oxford OX1 3RH, UK
\\
$^{6}$Departments of Physics and Astronomy, University of Illinois at Urbana-Champaign, 1002 W Green St, Urbana, IL 61801, USA\\
}

\date{Accepted XXX. Received YYY; in original form ZZZ}

\pubyear{2018}

\begin{document}
\label{firstpage}
\pagerange{\pageref{firstpage}--\pageref{lastpage}}
\maketitle


\begin{abstract}Although the Cosmic Microwave Background agrees with a perfect blackbody spectrum within the current experimental limits, it is expected to exhibit certain spectral distortions with  known spectral properties. We propose a new method,  Frequency Space Differential (FSD) to measure the spectral distortions in the CMB spectrum by using the inter-frequency differences of the brightness temperature. The difference between the observed CMB temperature at different frequencies must agree with the frequency  derivative of the blackbody spectrum, in the absence of any distortion. However, in the presence of spectral distortions, the measured inter-frequency differences would also exhibit deviations from blackbody which can be modeled for known sources of spectral distortions like $y\,\&\,\mu$. Our technique uses FSD information for the CMB blackbody, $y$, $\mu$ or any other sources of spectral distortions to model the observed signal. Successful application of this method in future CMB missions can provide an alternative method to  extract spectral distortion signals and can potentially make it feasible to measure spectral distortions without an internal blackbody calibrator.
\end{abstract}
\begin{keywords}
CMB spectral distortion, measurement technique
\end{keywords}

\section{Introduction}
Imprints of spectral distortions in the Cosmic Microwave Background (CMB) are a prediction of the Standard Cosmological Model \citep{zeldovich, Chluba:2011hw, Chluba:2012gq, Khatri:2012tv, Khatri:2012rt, Khatri:2012tw, Chluba:2016bvg, Hill:2015tqa, Emami:2015xqa}. Measurement of signals such as $y$ and $\mu$ distortions will help to validate our standard cosmological model. Indeed the essential ansatz of structure formation by gravitational instability predicts weak but potentially measurable $\mu$ distortions \citep{1994ApJ...430L...5H,2012PhRvL.109b1302P}. Discoveries of any other kinds of spectral distortions can open up a window to new physics. One of the main goals of several next generation cosmology missions is to measure the spectral distortions in the CMB blackbody spectrum. The first observational bound on the spectral distortion was given by FIRAS \citep{Mather:1993ij,firas,cobe} with $\mu < 9\times 10^{-5}$ and  $ y < 15\times 10^{-6}$ at $95\%$ C.L. FIRAS used an absolute blackbody internal calibrator to measure the monopole of the CMB temperature field and constrained its temperature  $T=2.725 \pm 0.001$ K \citep{Mather:1998gm, 2009ApJ...707..916F}. With the recent CMB anisotropy data, measurements of Sunyaev-Zel'dovich (SZ) clusters \citep{Hasselfield:2013wf, Bleem:2014iim, Ade:2015gva, Staniszewski:2008ma} and  bounds on the fluctuating $y$ and $\mu$ have also been obtained \citep{Khatri:2015jxa, Khatri:2015tla}.

Several concepts are under discussion for a post-Planck CMB polarization mission in space, including  spectrometry (PIXIE \citep{2011JCAP...07..025K} or PRISM \citep{Andre:2013nfa}), high-resolution imaging (CMBPOL) \citep{Dunkley:2008am}, or a mission with modest resolution  focusing on the large-angle primordial anisotropy (LiteBIRD) \citep{Matsumura:2016sri}. While imaging and spectroscopy are often presented as mutually exclusive concepts we propose  a hybrid approach to image spectral distortions which we term the Frequency Space Differential (FSD) method. 
This technique uses a differential measurement of the CMB between different frequencies and therefore  does not require an absolute calibrator. 

In this paper, we propose a new technique for measuring the spectral distortions in the CMB which can avoid using an absolute blackbody calibrator.
 We describe the possibility of measuring the spectral distortion in CMB by measuring the inter-frequency differences of the sky intensity and matching it with the theoretical prediction of the signal frequency spectrum. The blackbody spectrum predicts a well-known intensity or brightness temperature 
at every frequency. As a result, the difference of the blackbody intensity between two different frequencies can predict a unique spectral shape, and we can obtain an all-sky frequency space derivative map of the blackbody spectrum. In the presence of any spectral distortions in the blackbody  intensity, the derivative of the observed intensity is a composite signature of frequency derivatives of the blackbody spectrum and other sources of spectral distortions. 

We will see that the key idea is to design the measurement such that any gain fluctuations couple only to the frequency derivative of the blackbody spectrum rather than the blackbody spectrum itself.  A similar approach was  discussed recently by \cite{Sironi:2016dsh} who also proposed a detector design. 

Even in the absence of an absolute calibrator, an overall calibration can be obtained from the time-dependent  velocity dipole due to the orbital motion around the sun. This effect can be extracted from a multi-year campaign due to its annual modulation. It's frequency spectrum only depends on well-known relativistic effects and is directly proportional to the derivative of the blackbody spectrum. It  can therefore serve as an absolute and robust calibrator for the FSD technique. This effect was used in \citet{2009ApJ...707..916F} to recalibrate the FIRAS data using the WMAP time ordered data.

In this paper we discuss the main idea of using the frequency derivative of the spectral distortion signal to measure the $\mu$ and $y$ distortions without an absolute calibrator (in an analogous way to how   WMAP created a map from a purely differential measurement of the anisotropies, in contrast to Planck which used an internal reference).
 We prescribe possible measurement strategies and statistical techniques to implement this method for future CMB missions. The implementation of this method to a particular mission puts requirements on measurement technique, scan strategy, detector properties and calibration techniques. 
 
This paper is organized as follows: Sec.~\ref{formalism-main} sets out the form of the expected signals in the sky when observed differentially in frequency space. In Sec.~\ref{sys} we discuss how to form differential combinations of nearby frequency channels such that inter-channel calibration errors do not couple to the CMB monopole but only to the derivative of the Planck spectrum. Once this major source of noise is removed, the remaining signal needs to be cleaned from foreground contamination. A method for removal of those contaminants and recovery of the spectral distortion signal is given in Sec.~\ref{milc}. In Sec.~\ref{requirement} we discuss the main requirements our approach places on instrument design. We conclude in Sec.~\ref{conclusions}.

\section{Formalism}\label{formalism-main}
The all-sky average temperature field of the CMB exhibits a blackbody spectrum ($S_{\text{brightness}}\equiv c^2B_\nu/2k_B\nu^2= h\nu/k_B(e^{h\nu/k_BT_{\text{CMB}}}$-1) with a brightness temperature $S_{\text{brightness}}= 2.7255$ K in the RJ limit ($h \nu/k_B T_{\text{CMB}}<1$). Any deviation from the blackbody spectrum can be parametrised as \citep{Mather:1993ij,firas}
\begin{align}\label{source-f1}
I^o_\nu= B_\nu (T_{\text{CMB}}) + \Delta T_{CMB} \frac{\partial B}{\partial T} +\Delta I^{gal}_\nu + u \frac{\partial B}{\partial u},
\end{align}
where $I^o_\nu$ is the observed intensity in the sky and the first and second terms are the blackbody spectrum and fluctuations in the blackbody due to CMB temperature fluctuations.  The third term indicates the galactic contamination and the last term is the spectral distortion due to cosmological processes (like $u  \equiv\mu,\,y$). The observed intensity of the sky at every frequency should be compared with an internal blackbody calibrator fixed at a particular temperature to deduce the temperature of the CMB field and also any departure from blackbody. 
The FIRAS \citep{Mather:1993ij,firas} experiment used an internal blackbody calibrator to measure the CMB temperature field and also provided the first observational constraints on $\mu,\, y$ distortions as $9\times 10^{-5}$ and $15\times 10^{-6}$ at $95\%$ C.L. respectively.  Measurement of any well-motivated CMB spectral distortions to values of cosmological interest \citep{zeldovich, 1994ApJ...430L...5H,2010MNRAS.402.1195C, 2012MNRAS.419.1294C, Chluba:2011hw, Chluba:2012gq, Khatri:2012tv, Khatri:2012rt, Khatri:2012tw, Chluba:2016bvg, Hill:2015tqa, Emami:2015xqa} requires a much better absolute blackbody calibrator than FIRAS.  

We will show how to estimate cosmological spectral distortions with any given spectrum using the FSD technique. Astrophysical sources add contaminations with approximately known spectra. We will find that these have a similar effect on the FSD technique as on absolutely calibrated spectral distortion measurements. 


\subsection{Probing spectral distortions through spectral derivatives}\label{formalism-sub}
The observed CMB blackbody intensity, along with  spectral distortions like $y\, \&\, \mu$, also gets contaminated by several galactic astrophysical emissions in the  CMB frequency range by processes like synchrotron, free-free, spinning dust, thermal dust, etc. The total emission can be written in terms of a brightness temperature at a particular frequency $\nu$ in a particular pixel ($\hat p$) as a superimposition of various effects which can be written in the form
\begin{align}\label{source-1}
\begin{split}
S_\nu(\hat p)= &K^{\text{pl}}_{\nu} + A_{\text{CMB}}(\hat p) K^T_\nu + A_\mu(\hat p) K^\mu_\nu + A_y(\hat p) K^y_\nu  \\ & + A_{\text{dust}}(\hat p) K^{\text{dust}}_\nu + A_{\text{syn}}(\hat p) K^{\text{syn}}_\nu + A_{\text{free}}(\hat p) K^{\text{free}}_\nu  \\ & + A_{\text{spin-dust}}(\hat p) K^{\text{spin-dust}}_\nu,
\end{split}
\end{align}
where with $x= h \nu/k_BT_{CMB}= \nu/\nu_{CMB}$, we can write \citep{Adam:2015wua}
\begin{align}\label{source-1a}
\begin{split}
K^{\text{pl}}_{\nu} &= \frac{xT_{CMB}}{(e^x-1)},\\
K^T_\nu&= \frac{x^2e^x}{(e^x-1)^2},\\
K^\mu_\nu&= \frac{-xe^xT_{CMB}}{(e^x-1)^2},\\
K^y_\nu&= \frac{x^2e^xT_{CMB}}{(e^x-1)^2}\bigg(x\big(\frac{e^x+1}{e^x-1}\big)-4\bigg),\\
K^{\text{Dust}}_{\nu}&= \bigg(\frac{\nu}{\nu_0}\bigg)^{\beta_d+1}\bigg(\frac{e^{\nu_0/T_d}-1}{e^{\nu/T_d}-1}\bigg), \hspace{0.5cm} \\ &T_d=18 \text{K}, \nu_0= 545 \text{GHz}, \beta_d=1.55,\\
K^{\text{syn}}_{\nu}&= \bigg(\frac{\nu_0}{\nu}\bigg)^2 \frac{f_s(\nu/\alpha)}{f_s(\nu_0/\alpha)}\hspace{0.5cm}  \nu_0= 408 \text{MHz}, f_s= \text{templates},\\
K^{\text{free}}_\nu&= T_e(1-e^\tau), \hspace{0.5cm} \\&\tau= 0.05468 T_e^{-3/2}\nu_9^{-2} \log(e^{[5.96- \sqrt{3}{\pi}log(\nu_9T_4^{-3/2})]}+e), \\& \nu_9= \frac{\nu}{GHz}, T_4= T_e/10^4,\\
K^{\text{spin-dust}}_{\nu}&= \bigg(\frac{\nu_0}{\nu}\bigg)^2 \frac{f_{sd}(\nu.\nu_{p0}/\nu_p)}{f_{sd}(\nu_0.\nu_{p0}/\nu_p)}\hspace{0.5cm} \nu_{p0}=30 \text{GHz}, f_{sd}= \text{templates}.
\end{split}
\end{align}

The all-sky average measurement of Eq. \eqref{source-1} obtains the contribution only from the monopole term, whereas  the differential measurement of  Eq. \eqref{source-1} between different pixels gets the contribution only from the fluctuation parts ($\Delta T_{CMB} (\hat p), \Delta \mu (\hat p)$ and $\Delta y (\hat p)$), and captures no contributions from the monopole of the CMB or any other spectral distortion signals. However, the differential measurement  in  frequency space $\mathcal{S}_{\nu_{ji}}= S_{\nu_j}-S_{\nu_i}$ has  non-zero contributions both from the monopole and from the fluctuation part,  which can be written as 
\begin{align}\label{source-1a}
\begin{split}
\mathcal{S}_{\nu_{ji}}(\hat p)= &\mathcal{K}^{\text{pl}}_{\nu_{ji}} + A_{\text{CMB}}(\hat p) \mathcal{K}^T_{\nu_{ji}} + A_\mu(\hat p) \mathcal{K}^\mu_{\nu_{ji}} \\ &  + A_y(\hat p) \mathcal{K}^y_{\nu_{ji}}  + A_{\text{dust}}(\hat p) \mathcal{K}^{\text{dust}}_{\nu_{ji}} + A_{\text{syn}}(\hat p) \mathcal{K}^{\text{syn}}_{\nu_{ji}}\\ & + A_{\text{free}}(\hat p) \mathcal{K}^{\text{free}}_{\nu_{ji}}  + A_{\text{spin-dust}}(\hat p) \mathcal{K}^{\text{spin-dust}}_{\nu_{ji}}.
\end{split}
\end{align}
For closely spaced frequency channels, i.e. small $\Delta \nu_{ji} = \nu_j-\nu_i$ this can be related to the derivative of the theoretical frequency spectrum $\mathcal{K}^{x}_{\nu_{ji}}(\hat p)= \partial K^x/\partial \nu|_{\nu_{ji}} \Delta \nu_{ji}$ evaluated at the midpoint $\nu_{ji}= (\nu_j+\nu_i)/2$.  The theoretical Frequency Space Derivative (FSD) spectrum of the various sources can be expressed in terms of  $\mathcal{D}^x_{\nu_0}(\hat p)\equiv (\partial K^x/\partial \nu)|_{\nu_0}$  as
\begin{align}\label{source-2a}
\begin{split}
\mathcal{D}^{\text{pl}}_{\nu} &= \frac{A_{CMB}}{\nu_{CMB}}\frac{1}{(e^x-1)}\bigg[1-\frac{xe^x}{(e^x-1)}\bigg],\\
\mathcal{D}^T_\nu&= \frac{1}{\nu_{CMB}}\frac{x^2e^x}{(e^x-1)^2}\bigg[\frac{2}{x}+1-\frac{2e^x}{(e^x-1)}\bigg],\\
\mathcal{D}^\mu_\nu&= \frac{T_{CMB}}{\nu_{CMB}}\frac{e^x}{(e^x-1)^2}\bigg[-1 -x + \frac{2xe^x}{(e^x-1)}\bigg],\\
\mathcal{D}^y_\nu&= \frac{T_{CMB}}{\nu_{\text{CMB}}}\bigg[\Delta n_\nu^y\bigg(2 + x-\frac{2xe^x}{(e^x-1)}\bigg) \\& + \frac{x^2e^x}{(e^x-1)^2}\bigg(\bigg(\frac{e^x+1}{e^x-1}\bigg) + \bigg(\frac{xe^x}{e^x-1}\bigg) - \bigg(\frac{xe^x(e^x+1)}{(e^x-1)^2}\bigg)\bigg)\bigg],\\
\mathcal{D}^{\text{Dust}}_{\nu}&= \bigg(\frac{\nu}{\nu_0}\bigg)^{\beta_d+1}\bigg(\frac{e^{\nu_0/T_d}-1}{e^{\nu/T_d}-1}\bigg)\bigg[\frac{\beta_d+1}{\nu} - \frac{\gamma e^{\gamma \nu}}{(e^{\gamma \nu}-1)}\bigg],\\
\mathcal{D}^{\text{syn}}_{\nu}&= \bigg(\frac{\nu_0}{\nu}\bigg)^2 \bigg(\frac{-2}{\nu}\frac{f_s(\nu/\alpha)}{f_s(\nu_0/\alpha)} + \frac{\partial f_s(\nu/\alpha)}{\partial \nu}\frac{1}{f_s(\nu_0/\alpha)}\bigg),\\
\mathcal{D}^{\text{free}}_\nu&= -T_ee^\tau \frac{\partial \tau}{\partial \nu},\\
\mathcal{D}^{\text{spin-dust}}_{\nu}&= \bigg(\frac{\nu_0}{\nu}\bigg)^2 \bigg(\frac{-2}{\nu}\frac{f_{sd}(\nu.\nu_{p0}/\nu_p)}{f_{sd}(\nu_0.\nu_{p0}/\nu_p)} \\& +  \frac{\partial f_{sd}(\nu.\nu_{p0}/\nu_p)}{\partial \nu}\frac{1}{f_{sd}(\nu_0.\nu_{p0}/\nu_p)}\bigg),
\end{split}
\end{align}
\begin{figure}
\includegraphics[width=3.5in,keepaspectratio=true]{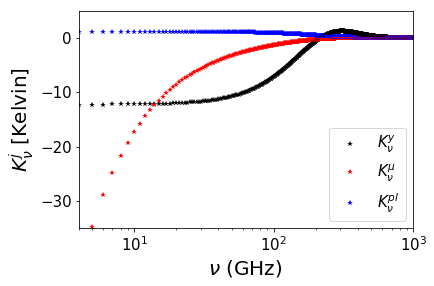}
\caption{We plot the kernel of the signal for different sources of spectral distortions for $y$-distortions (black) which is multiplied by a factor of  two in the amplitude, $\mu$-distortions (red) and blackbody spectrum (blue) at $T_0=2.7255$ Kelvin.}\label{cosmo-spectrum}
\end{figure}
\begin{figure}
\includegraphics[width=3.5in,keepaspectratio=true]{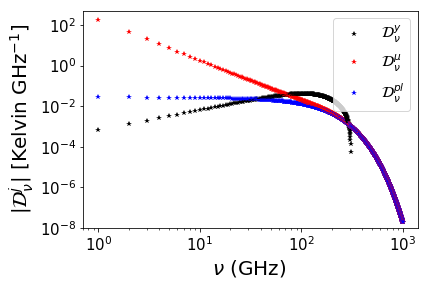}
\caption{The Frequency Space Derivative (FSD) spectrum of the $y$-distortions (black), $\mu$-distortions (red) and $ \text{blackbody}$ spectrum (blue) are depicted over  a wide frequency range which is usually accessible by CMB missions.}\label{cosmo-spectrum}
\end{figure}

where $\Delta n_\nu^y= K^y_{\nu}/xT_{CMB}$.
 Using Eq. \eqref{source-2a}, we can write Eq. \eqref{source-1a} as
\begin{align}\label{source-1b}
\begin{split}
\mathcal{S}_{\nu}(\hat p)= &\mathcal{D}^{\text{pl}}_{\nu}\Delta \nu + A_{\text{CMB}}(\hat p) \mathcal{D}^T_{\nu}\Delta \nu + A_\mu(\hat p) \mathcal{D}^\mu_{\nu}\Delta \nu \\& + A_y(\hat p) \mathcal{D}^y_{\nu}\Delta \nu +  A_{\text{dust}}(\hat p) \mathcal{D}^{\text{dust}}_{\nu}\Delta \nu \\&+ A_{\text{syn}}(\hat p) \mathcal{D}^{\text{syn}}_{\nu}\Delta \nu + A_{\text{free}}(\hat p) \mathcal{D}^{\text{free}}_{\nu}\Delta \nu \\&+ A_{\text{spin-dust}}(\hat p) \mathcal{D}^{\text{spin-dust}}_{\nu}\Delta \nu,\\
\end{split}
\end{align}
or in matrix notation
\begin{equation}
\mathbf{\mathcal{S}}(\hat p)= \mathcal{\mathbf{D}} \mathbf{A}(\hat p).
\end{equation}
Here $\mathcal{\mathbf{D}}$ is the matrix of the FSD spectrum with components $\mathcal{D}_{ji}= \mathcal{D}^{\text{j}}_{\nu_i}\Delta \nu_i$ and $\mathbf{A}$ is the column matrix composed of the signals. This equation relates the Frequency Space Differential (FSD) with the known theoretical spectrum of several sources. The FSD spectrum for different sources are plotted in Fig.~\ref{cosmo-spectrum}. As  is clear from Fig.~\ref{cosmo-spectrum}, the spectrum for each of the sources is  distinct.  For $\mu$ distortions, the FSD signal is mainly strong at low frequencies and  decays rapidly. The $y$ distortions peak at higher frequencies with a much wider FSD spectrum ($\nu \in 40-300$ GHz) than $\mu$. A mission to constrain both $y$ and $\mu$ therefore requires a combination of low and high frequency channels. We will leave a detailed design study of an optimal distribution of channel frequencies and bandwidths to future work.
An estimator such as Modified Internal Linear Combination \citep{Hurier2013} combines all  frequency channels to reject foreground contamination and improve  Signal to Noise  (SNR). We will develop the formalism of such an estimator in the context of the FSD technique in Sec.~\ref{milc}.

In the next section, we describe the dominant source of systematic error in this method and how to mitigate it.

\section{Measurement technique and systematic errors}\label{sys}
Usage of the FSD technique to measure CMB spectral distortions is only possible if the temperature differences between  frequency channels can be determined with sufficient  systematic error control.
Though the measurement techniques and detector properties depend on specific missions, we  discuss the basic requirements which should be addressed in order to use the FSD method in this section.

\subsection{The non-differential technique of measuring the radiation field}\label{non-diff}
To illustrate the problem we will first discuss why non-differential methods will not be able to provide useful constraints on spectral distortions. The radiation impinging on a pixel ($\hat p$) at a frequency ${\nu_i}$ of the detector produces a voltage $V_{\nu_i}(\hat p)$ which is related to the observed temperature field by the gain factor $G_{\nu_i}$ as 
 \begin{equation}\label{sys-1}
T_{\nu_i}(\hat p)= G_{\nu_i} V_{\nu_i}(\hat p) + T^{\text{off}}_{\nu_i},
\end{equation}
where $T^{\text{off}}_{\nu_i}$ is the instrumental off-set temperature.  In the absence of mean detector noise, the measured temperature at a pixel $\hat p$ in  frequency $\nu_i$ is related to the theoretical FSD signal by the relation,
 \begin{equation}\label{theo-1}
T_{\nu_i}(\hat p) = S_{\nu_i}(\hat p) T^{\text{off}}_{\nu_i}.
\end{equation}
However, detectors even with a known and stable gain factor $G_{\nu_i}$ and offset temperature $T^{\text{off}}_{\nu_i} =0$, exhibit variations $\delta G_{\nu_i}$ and $\delta T^{\text{off}}_{\nu_i}$ which are the sources of systematic errors that propagate through the measurements. As a result, the systematic error associated with the temperature field (Eq. \eqref{sys-1}) are due to the gain and off-set error, which can be written as  
 \begin{equation}\label{sys-1a}
(\sigma_{\nu_i})^2_{\text{sys}} \equiv (\delta T_{\nu_i}(\hat p))^2= \bigg(\frac{\delta G_{\nu_i}}{G_{\nu_i}}\bigg)^2 (G_{\nu_i}V_{\nu_i}(\hat p))^2 + (\delta T^{\text{off}}_{\nu_i})^2.
\end{equation}
After using Eq. \eqref{theo-1}, above equation can be expressed as
\begin{equation}\label{systh-1a}
(\sigma_{\nu_i})^2_{\text{sys}} = \bigg(\frac{\delta G_{\nu_i}}{G_{\nu_i}}\bigg)^2 (S_{\nu_i}(\hat p))^2 + (\delta T^{\text{off}}_{\nu_i})^2.
\end{equation}
The dominant contribution to $S_{\nu_i}(\hat p)$ is the blackbody temperature field of CMB ($K^{pl}_{\nu_i}$ in Eq. \eqref{source-1}). As a result, the dominant source of systematic error in Eq. \eqref{systh-1a} is induced by the coupling between gain error $\delta G_{\nu_i}$ and $K^{pl}_{\nu_i}$. For the typical values of gain error (of order $0.1\%-0.01\%$) \footnote{A detailed description of the systematic error is given in Sec. \ref{sys-error}}, the contribution of the systematic error is greater than the usual signal strength of $\mu$ and $y$ distortions. So to measure the spectral distortion signals, an absolute internal blackbody calibrator with a precisely known reference temperature is required under this method of measurement. 

In the following subsection, we will introduce a \textit{new differential method} which can reduce the systematic error without an absolute calibrator and also use the cross-calibration between the frequency channels to minimize the budget of the systematic error.

\subsection{The FSD technique for measuring the radiation field}\label{fsdm}
The incoming electromagnetic waves (composed of multiple components) from the sky at a particular frequency channel $\nu_i$ falls on the detector (operating at this frequency) will have an induced voltage, which we  define as $V_{\nu_i}$. 
In the FSD technique, we propose the measurement of the difference in the amplitude of electromagnetic field at two different frequencies by taking the difference between the induced voltages ($V_{\nu_j}- V_{\nu_i}$). 
This differential measurement of the signal carries the information of the change in the electromagnetic field of CMB (and also in other contaminations) with variation in the frequency. The measured differential voltage can be converted into a temperature difference by a known gain factor $G_{\nu_i}$ and $G_{\nu_j}$ by the relation
\begin{equation}\label{sys-5}
\mathcal{T}_{\nu_{ji}}(\hat p)= G_{\nu_j}\delta V_{\nu_{ji}}(\hat p) + G_{\nu_i}\delta V_{\nu_{ji}}(\hat p), 
\end{equation}
where, $\delta V_{\nu_{ji}}(\hat p)= (V_{\nu_j} - V_{\nu_i})/2$, $\nu_{ji}= (\nu_j +\nu_i)/2$ and the off-set temperature difference between the two channels is assumed to be zero. If $G_{\nu_i} = G_{\nu_j}$, then the above equation is directly related to the theoretical FSD signal $\mathcal{S}_{\nu_{ji}}$ (Eq. \eqref{source-1b}) as
\begin{equation}\label{systheo-5}
\mathcal{T}_{\nu_{ji}}(\hat p)\equiv G_{\nu_j}(\delta V_{\nu_{ji}}(\hat p) + \delta V_{\nu_{ji}}(\hat p)) = \mathcal{S}_{\nu_{ji}}(\hat p). 
\end{equation}
However, if  $G_{\nu_i} = G_{\nu_j} +\Delta G_{\nu_{ji}}$, then 
 \begin{equation}\label{systheo-5a}
\mathcal{T}_{\nu_{ji}}(\hat p) = \mathcal{S}_{\nu_{ji}}(\hat p) + \frac{\Delta G}{G_{\nu_j}} G_{\nu_j}\delta V_{\nu_{ji}}(\hat p), 
\end{equation}
where, the second term is an extra bias originating from the difference of the gain factors between two frequency channels.  This indicates that any variation in gain factor will affect the measurement by coupling it with the difference in the voltages and not with the absolute value of the voltages. As the voltage difference between two channels have the dominant contribution from the FSD spectrum of the blackbody $\mathcal{S}^{pl}_{\nu_{ji}}$, the above equation can be approximated as
 \begin{equation}\label{systheo-5b}
\mathcal{T}_{\nu_{ji}}(\hat p) \approx \mathcal{S}_{\nu_{ji}}(\hat p)+ \frac{\Delta G}{G_{\nu_j}}\mathcal{S}^{pl}_{\nu_{ji}}(\hat p). 
\end{equation}

The variance of Eq. \ref{sys-5}, only due to the uncorrelated systematic errors in the gain factor and off-set temperature can be written as
\begin{align}\label{sys-6}
\begin{split}
(\sigma^2_{\nu_{ji}})_{sys}\equiv(\delta \mathcal{T}_{\nu_{ji}})^2=& \bigg(\frac{\delta G_{\nu_i}}{G_{\nu_i}}\bigg)^2 G^2_{\nu_i}\delta V^2_{\nu_{ji}} + \bigg(\frac{\delta G_{\nu_j}}{G_{\nu_j}}\bigg)^2 G^2_{\nu_j}\delta V^2_{\nu_{ji}} \\&+ (\delta T^{\text{off}}_{\nu_j})^2 + (\delta T^{\text{off}}_{\nu_i})^2.
\end{split}
\end{align}
Here, $(\delta T^{\text{off}}_{\nu_i})^2$ denotes the variance in the offset measurement. The systematic error is related to the voltage difference which according to Eq. \eqref{systheo-5a}, have the major contribution from the FSD spectrum of CMB blackbody. 

The comparison of Eq. \eqref{sys-1a} and Eq. \eqref{sys-6} exhibits the key difference between the non-differential technique and the FSD technique. The systematic error is related to the absolute blackbody signal in the former case and to the derivative of the blackbody in the latter case. As depicted in Fig.~\ref{cosmo-spectrum}, for $\Delta \nu=1$ GHz, the amplitude of FSD spectrum of blackbody is two orders of magnitude below the blackbody signal. Hence, the systematic error between these two methods will also differ by two orders of magnitude. 

The total error due to both the systematic and the statistical error can be written as
\begin{equation}\label{sys-7b}
(\sigma^2_{\nu_{ji}})_{tot}= (\sigma^2_{\nu_{ji}})_{sys}+ (\sigma^2_{\nu_{ji}})_{stat},
\end{equation}
where, we define the statistical error in terms of the uncorrelated instrumental noise as
\begin{align}\label{sys-7}
\begin{split}
(\sigma^2_{\nu_{ji}})_{stat} (\hat p) &=(\delta T^N_{\nu_{i}}(\hat p))^2 + (\delta T^N_{\nu_{j}}(\hat p))^2.
\end{split}
\end{align}
 
\subsection{Required optimization for a multi-frequency system}
The above-mentioned FSD technique is a differential measurement of the imaging signal obtained from the high resolution frequency bands to construct the deviations from blackbody. Implementation of this method along with the standard imaging method (by using low resolution frequency bands) is required to achieve the science goals from the spectral distortions as well as the anisotropic part of CMB. So we need a  hybrid composition of frequency resolution to implement both FSD technique and imaging technique, such that we can obtain the spectral distortion signal \& anisotropic signal from the same conceptual framework and also with minimum cost and minimum error. 

To minimize the sources of systematic error, it is required to reduce the contribution of CMB blackbody in the differential measurement between two different frequency channels. So we need high resolution frequency channels to subtract the blackbody part substantially so that the total systematic error is smaller than the spectral distortion signal. The high resolution FSD technique needs to be implemented on the frequency range which have the large values of FSD kernel for $\mu$ distortion (approximately $ 1-50$ GHz) and $y$ distortion (approximately $100-300$ GHz), which can be identified from Fig.~\ref{cosmo-spectrum}. The  remaining frequency ranges can have large bandwidth to perform  the scientific studies related to imaging. A detailed case study of the FSD technique can be done for a specific mission with the knowledge of the detector properties, calibration error, data read-out frequency, size of the focal plane etc.

For multiple frequency channels, we require to estimate the covariance matrix consisting of contributions from systematic errors\footnote{Bold fonts denotes matrices}  ($\textbf{C}_{\text{sys}}$), statistical error $(\textbf{C}_N)$ and error due to each cosmological and astrophysical component ($\textbf{C}_A \equiv \langle \mathbf{A} \mathbf{A}^\dagger \rangle$). 
So the total covariance matrix becomes

\begin{align}\label{sys-8a}
\langle \mathcal{T} \mathcal{T}^\dagger \rangle \equiv  \textbf{C}_\mathcal{T}&= \textbf{D}\textbf{C}_A\textbf{D}^T + \textbf{C}_{\text{sys}} + \textbf{C}_N.
\end{align}
The covariance matrix is not diagonal and needs to be evaluated for every mission with the particular instrumental noise, systematic errors and frequency coverage. The essential requirement to implement the FSD technique is to reduce the contribution of the total error on the signals of spectral distortion. The total contribution from the systematic and the instrumental noise matrix can be written as
\begin{align}\label{sys-8b}
\textbf{C}_\mathcal{\tilde N}&= \textbf{C}_{\text{sys}} + \textbf{C}_N, 
\end{align}
which can be decomposed as
\begin{align}\label{sys-8c}
\textbf{C}_\mathcal{\tilde N}&= \textbf{E}\mathbf{\Gamma}\textbf{E}^{T},
\end{align}
where, \textbf{$\Gamma$} is a diagonal matrix of eigenvalues $\gamma_1, \gamma_2, \hdots ,\gamma_n$ such that $\gamma_1>\gamma_2>\hdots> \gamma_n$ and the matrix $\textbf{E}$ contains the corresponding eigenvectors. An experimental design which can achieve the condition that the eigenvectors with largest eigenvalues have a minimum projection on the FSD kernel (like $D_\nu^\mu$ and $D_\nu^y$ for $\mu$ and $y$ respectively) can significantly improve the SNR of the measurement. 
In Sec. \ref{milc}, we elaborate more on this and also explain the procedures to extract the signal. 

\subsection{Gain factor and the calibration error}\label{sys-error}
For CMB experiments, there are several standard calibration sources   \citep{Adam:2015vua} like the CMB solar dipole, the orbital dipole and planets. These  are used for calibration by the  Planck mission \citep{Adam:2015vua}. The orbital dipole is a very good calibrator due to the well-known value of the satellite velocity and gives a very small calibration error of typically $0.1\%-0.01\%$ \citep{Adam:2015vua}. The motion of the solar barycentre (dipole) or the orbital motion (of the satellite) also exhibits a known spectrum which can be written as $\Delta T_{dip}K^T_{\nu_i}$, where $\Delta T_{dip}$ is the magnitude of the induced temperature due to the solar or the orbital motion and $K^T_{\nu_i}$ is the derivative of the blackbody spectrum with respect to the temperature. So using the known value of the brightness temperature of the CMB dipole and (time-dependent) orbital motion, we can calibrate the detectors for each frequency channel. For the remaining discussion in the paper we will focus only on the orbital dipole because of several factors like (i) ease of modelling accurately, (ii) measurement with  very high SNR by current detectors and (iii) clean demodulation from multi-year data due to its annual variation.
 {Schematically, the CMB dipole can be measured within each frequency channel through its pixel-to-pixel variation
\begin{equation}\label{dip-1}
\delta S^{\text{dip}}_{\nu_i}\equiv S_{\nu_i}(\hat p_1) - S_{\nu_i}(\hat p_2)= G_{\nu_i}(V_{\nu_i}(\hat p_1) - V_{\nu_i}(\hat p_2))\equiv G_{\nu_i}\Delta V_{\nu_i},
\end{equation}
Due to the known frequency spectrum of $\delta S^{\text{dip}}_{\nu_i}$, one can write this as
\begin{equation}\label{dip-2}
G_{\nu_i}= \frac{K^T_{\nu_i} \Delta T_{\text{dip}}}{\Delta V_{\nu_i}}.
\end{equation}
As a result, the gain error in terms of the error associated with the dipole measurement ($\delta (\Delta T_{\text{dip}})$) and voltage measurement ($\delta (\Delta V)$) can be written as
\begin{equation}\label{dip-3}
\bigg(\frac{\delta G_{\nu_i}}{G_{\nu_i}}\bigg)^2\simeq \bigg(\frac{\delta (\Delta T_{\text{dip}})}{\Delta T_{\text{dip}}}\bigg)^2 + \bigg(\frac{\delta (\Delta V)}{\Delta V}\bigg)^2.
\end{equation}
So the gain error of each channel is related to the error associated with the measurement of dipole amplitude, even if the error in the measurement of voltage is negligible.}

We can accurately obtain the relative gain coefficients at different frequencies by cross-calibrating between frequency channels. By equating the dipole amplitude fluctuation $\Delta T_{\text{dip}}$ between any two frequency channels, we can write
\begin{equation}\label{dip-4}
\frac{G_{\nu_j} \Delta V_{\nu_j}}{K^T_{\nu_j}} = \frac{G_{\nu_i} \Delta V_{\nu_i}}{K^T_{\nu_i}},
\end{equation}
which implies 
\begin{equation}
\mathcal{G}_{ji} \equiv \frac{G_{\nu_j}}{G_{\nu_i}} = \frac{\Delta V_{\nu_i}}{\Delta V_{\nu_j}} \frac{K^T_{\nu_j}}{K^T_{\nu_i}}.
\end{equation}
This indicates that the relative calibration depends only on the measured voltage difference. Therefore, the corresponding error in the ratio of the gain is affected only by the error associated with the measurement of voltage difference and not that associated with the orbital dipole measurement. As a result, the error on the relative gain ratio can be reduced. This also indicates that the accurate calibration of the gain factor at any one frequency channel translates into an accurate calibration at all channels.

\section{Signal extraction using different techniques}\label{milc}
 \subsection{Fitting the FSD spectrum}
The FSD measurement of the all-sky intensity (or equivalently brightness temperature) at different frequency channels is an addition of several signals due to cosmological and astrophysical sources and also instrumental noise. With the known spectrum of the FSD and a high spectral resolution measurement over a wide frequency band, we can estimate the best-fit parameter $\mathcal{\hat A}_{x}$ (where $x \in [y,\, \mu,\hdots]$), which minimizes the chi-square defined as
\begin{equation}\label{extract-7}
\chi_{y\,,\mu}^2= \sum_{\nu, \nu'}\bigg(\bar{\mathcal{T_\nu}}- \mathcal{\hat A}_{y,\mu} \mathcal{D}^{y,\mu}_\nu\bigg) (C_{\mathcal{T}}^{-1})_{\nu\nu'} \bigg(\bar{\mathcal{T_{\nu'}}}- \mathcal{\hat A}_{y,\mu} \mathcal{D}^{y,\mu}_{\nu'}\bigg).
 \end{equation}
Addition over a wide range of frequencies increases the over-all SNR of the signal. The corresponding error bar on $ \mathcal{\hat A}_{y,\mu}$ is a standard result given by
\begin{equation}\label{extract-7a}
\sigma^2_{y,\mu}= \bigg[\sum_{\nu, \nu'} \mathcal{D}^{y,\mu}_{\nu} (C_{\mathcal{T}}^{-1})_{\nu\nu'} \mathcal{D}^{y,\mu}_{\nu'} \bigg]^{-1}.
 \end{equation}
As mentioned before, the covariance matrix $C_{\mathcal{T}}$ is non-diagonal and is a quantity which depends upon instrumental noise, scanning strategy, systematic errors, etc. For a particular mission, these quantities need to be evaluated for successfully implementing the FSD technique. 

\subsection{Internal Linear Combination Method}
After the removal of the coupling between gain errors and the CMB monopole the main remaining  hurdle to measuring $\mu$ \& $y$ distortions is  foreground contamination.  At low frequencies, the main sources of contamination are synchrotron emission and spinning dust emission from our galaxy.  At high frequency, foreground contamination is mainly due to dust. Since the FSD spectrum of $\mu$ and $y$ are not degenerate with these foregrounds we will now discuss how to use combinations of frequency channels over a wide range of frequencies to project out foreground contamination. 

 We at first address the extraction of the monopole part of the spectral distortion signal by an all-sky average of the FSD spectrum. 
The all-sky average value of the distortion signal can be extracted using the known FSD spectrum (Eq. \eqref{source-2a}) by the Internal Linear Combination (ILC) \citep{2011MNRAS.410.2481R} and Modified Linear Combination (MILC) \citep{Hurier2013}. With $M_\nu$ frequency channels over which $\mathcal{T}_{i}$ is estimated,  we can write
 \begin{equation}\label{extract-2}
 \mathcal{T}_{\nu_i} \equiv \mathcal{T}_{\nu_i} (\hat p)= \mathcal{D}^j_{\nu_i}\Delta \nu_i A^j(\hat p) + N_{\nu_i}(\hat p),
\end{equation}
where, $i \in[1, {M_\nu}]$ and $A^j \equiv [A_{CMB}, A_{\mu}, A_{y}, \hdots, A_{N_s-1}]$. $A^j$ contains both cosmological signal and also foreground contaminations.   
In terms of the $M_\nu \times 1$ column vector $\bf {\mathcal{T}}$, $N_s \times 1$ column vector $\bf {A}$ and $M_\nu \times N_s$ mixing kernel $\bf{D}$, we can write
\begin{equation}\label{extract-2a}
\bf{\mathcal{T}}= \bf{\mathcal{\mathbf{D}}}\bf{A} + \bf{N}.
\end{equation}
In the  presence of a non-zero value of $\Delta G$ (introduced in the previous section), there is also an additional component given by
 \begin{align}\label{extract-2b}
\mathcal{T}_{\nu_i}&= \mathcal{D}^j_{\nu_i}\Delta \nu_{i} A^j(\hat p) + N_{\nu_i} (\hat p) + \frac{\Delta G_{\nu_i}}{G_{\nu_i}}\mathcal{S}^{pl}_{\nu_i},\\
\bf{\mathcal{T}}&= \bf{\mathcal{\mathbf{D}}}\bf{A} + \bf{N} + \textbf{J}_G,
\end{align}
where, $\mathcal{S}^{pl}_{\nu_i}$ is the FSD blackbody spectrum at frequency $\nu_i$ and $\textbf{J}_G$ is the residual column matrix which can arise due to difference in the gain of the  frequency channels.
The extraction of the signal is achievable with the requirement that we recover only one component and follow the constraint that all other components do not contribute to the signal. For $N_s$ rejected components, we can define weights $\bf{w}$ such that 
\begin{align}\label{extract-3}
\begin{split}
u_1&= \mathbf{w^T f_1}=0,\\
u_2&= \mathbf{w^T f_2}=0,\\
&\vdots\hspace{1cm} \vdots \\
u_{j}&= \mathbf{w^T f_{j}}=1,\\
&\vdots\hspace{1cm} \vdots \\
u_{N_s}&= \mathbf{w^T f_{N_s}}=0,\\
u_{N_s+1}&= \mathbf{w^T J_G}=0,
\end{split}
\end{align}
where  {$\mathbf{f}_{j}$ are the frequency dependence of the $j^{th}$ signal defined as $\mathbf{f_j}= \mathbf{\mathcal{D} x_j}$}. $\mathbf{f}_{j}$ is a column vector with $M_\nu \times 1$ elements and $\mathbf{x_j}= [0,0,\hdots{},1,\hdots,0]^T$ with only $j^{th}$ element equal to one. 
The last condition of Eq. \eqref{extract-3} also put constraints on the nature of relative gain difference $\Delta G_i/G_i$. For the FSD technique to work, $J_G$ should not behave like any of the spectral signatures like $\mu$, $y$, etc. and hence needs to satisfy the condition 
 {
\begin{align}\label{extract-3a}
\begin{split}
\left[\mathbf{J_G}\mathbf{f}_j\right]_{i} =
\frac{\Delta G_{\nu_i}}{G_{\nu_i}}\mathcal{S}^{pl}_{\nu_i} f^j_{\nu_i} =0.
\end{split}
\end{align}}
 {As $S^{pl}_{\nu_i}$ is the known FSD spectrum of blackbody at frequency $\nu_i$, so the required frequency dependence of $\Delta G_{\nu_i}/G_{\nu_i}$ to minimize the residual contaminations in the signal is manifested by Eq. \eqref{extract-3a}. A special case with $\Delta G_{\nu_i}=0$ is a trivial solution of this equation and is sufficient but not necessary to be satisfied by the detectors. In the remaining of the paper, we will assume that the correction from $\Delta G_{\nu_i}/G_{\nu_i}$ can be made and we restrict only to $N_s$ values of $u$.}

With the requirement that the variance in the extracted  \textit{signal} map $\mathbf{C}_\mathcal{\hat{A}} = \langle \hat A \hat A^T\rangle$ is minimum, the weight matrix can be obtained  by solving the equation 
{\begin{equation}\label{extract-5}
\begin{bmatrix}
    2\mathbf{C_{\mathcal{T}}} & -\mathbf{D} \\
     {\mathbf{D}^T} & 0
  \end{bmatrix}
   \begin{bmatrix}
    \mathbf{w} \\
    \mathbf{\lambda}
  \end{bmatrix}
  =  \begin{bmatrix}
    0 \\
    \mathbf{x}
  \end{bmatrix},
 \end{equation}
where $\mathbf{\lambda}$ is the Lagrange multiplier and the covariance matrix $\mathcal{\mathbf{C}_{\mathcal{T}}}$ is a $M_\nu \times M_\nu$ dimension matrix which can be expressed as
\begin{equation}\label{extract-4b}
\textbf{$\mathcal{{\mathbf{C}_{\mathcal{T}}}}$}= \mathbf{D^TC_AD}+ \mathbf{C_{\bar{N}}},\\
\end{equation}
where $\mathbf{C_{\bar{N}}}$ is defined in Eq.\eqref{sys-8b} and have the contributions from instrumental noise, systematic errors and covariance matrix of the cosmological and astrophysical sources. 

The weight matrix which satisfies Eq. \eqref{extract-5} can be expressed as
\begin{equation}\label{extract-6}
\mathbf{W}=\mathbf{C^{-1}_\mathcal{T}} \mathbf{D}(\mathbf{D}^T\mathbf{C^{-1}_\mathcal{T}}\mathbf{D})^{-1},
 \end{equation}
and the corresponding $j^{th}$ component of the map can be obtained as 
\begin{equation}\label{extract-4}
\mathcal{\hat{A}}_j= \mathbf{x_j^TW^T\mathcal{T}}.
\end{equation}}
 The error estimate of the signal map $\mathcal{\hat {R}}_{j}$ can be written as
\begin{equation}\label{extract-4a}
\mathcal{{\mathbf{C}_{\mathcal{\hat A_{\text{j}}}}}}= \mathbf{x_j^TW^T\mathcal{\mathbf{C}_{\mathcal{T}}}Wx_j}.
\end{equation}
 
 {Using the above formalism for every component of the signal, we can obtain the weight matrix $W$ which minimizes the variance of the signal. To further reduce the error of the signal, we can satisfy the condition similar to Eq. \eqref{sys-8c} for the covariance matrix $\mathbf{C}_\mathcal{T}$ such that weight matrix projects minimally with the eigenvector corresponding to the largest eigenvalue of the covariance matrix. So the error estimate on the $j^{th}$ component in terms of the eigenvector decomposition ($\mathbf{C}_\mathcal{\tilde T}= \mathbf{E_\mathcal{\tilde T}}\mathbf{\Gamma_\mathcal{\tilde T}}\mathbf{E_\mathcal{\tilde T}}^{-1}$) can be written as
\begin{align}\label{eigen-1a}
\mathcal{{\mathbf{C}_{\mathcal{\hat A_{\text{j}}}}}}= \mathbf{x_j^TW^T\mathbf{E_\mathcal{\tilde T}}\textbf{$\Gamma_\mathcal{\tilde T}$}\mathbf{E_\mathcal{\tilde T}}^{-1}Wx_j},
\end{align}
which satisfies the condition 
\begin{align}\label{eigen-2a}
\mathbf{x_j^TW^T\mathbf{E^i_\mathcal{\tilde T}}} \approx 0\,\, \forall\, \gamma_i > \gamma_{\text{min}}.
\end{align}
}
where, $\gamma_{\text{min}}$ is the smallest eigenvalue of the covariance matrix $\mathcal{\mathbf{C}_{\mathcal{T}}}$.
\subsection{Measurement of spatial variations in the spectral distortion}\label{spatial}
The methods described previously have a particular application for approaching the monopole part of the spectral distortion signal. However this approach can be readily extendable to measure the fluctuations in the spectral distortion signal. Measurement of the FSD signal at every frequency channel gives a pixel space map of the signal, which in general can be written as
\begin{align}\label{spatial-1}
\mathcal{T}_{\nu_i}(\hat p)&= \sum_j D^{j}_{\nu_i} \Delta\nu_iA^j(\hat p),\\
(\mathcal{T}_{\nu_i})_{lm}&= \sum_{lm}\sum_j D^j_{\nu_i} \Delta\nu_i(A^j)_{lm},
 \end{align}
  {where, $\mathcal{T}_{\nu_i})_{lm}$ and $(A^j)_{lm}$ are the spherical harmonics transformation of $\mathcal{T}_{\nu_i}(\hat p)$ and $A^j(\hat p)$ respectively.}
The fluctuations in the signal can be captured by the power spectrum $_{\nu}C^{\mathcal{T}\mathcal{T}}_l= \frac{\sum_m (\mathcal{T}_{\nu})_{lm}(\mathcal{T}^*_{\nu})_{lm}}{(2l+1)}$ which is a composite effect of all the mechanisms. 
The dominant source of fluctuations in the spectral distortion is due to the y-distortion \citep{Hill:2015tqa}. With this technique, we can access the spatial fluctuations in the spectral distortions which are expected to be stronger than $\mu$ distortions. The intrinsic temperature fluctuations exhibit a  very different FSD spectrum from $y$ and hence are easily separable. The ILC method for FSD signal discussed previously is also directly applicable to reconstructing  the signal at every pixel and to generating a map of the fluctuations.

\section{Requirements to use FSD}\label{requirement}
While not using an absolute internal calibrator to measure CMB spectral distortions has clear practical advantages, the both approaches have their unique features and challenges.
For an absolute internal calibrator, it is essential that the calibrator is stable in temperature and is  a perfect blackbody so that it matches the blackbody distribution of CMB. Even a tiny departure from the blackbody spectrum of the absolute internal calibrator can    act as a source of systematic error and obscure any cosmological spectral distortion. 
In the absence of an absolute internal calibrator we have to achieve good control of the systematic errors in the temperature measurement and excellent relative calibration of different frequency channels, but for a potentially significant reduction in mission complexity and hence cost. 

We now discuss the necessary requirements  to use FSD for detecting CMB spectral distortions.  
\begin{enumerate}
\item{Instrument and measurement technique should be designed such that the final output is calibrated only with the relative voltage difference between two frequency channels as discussed in Sec. \ref{fsdm}.}
\item{The FSD signal due to $y$ and $\mu$ distortions peak at different frequency ranges as shown in Fig.~\ref{cosmo-spectrum}. So,  multiple high spectral resolution channels in those frequency range should be implemented with minimum instrumental noise. Use of high spectral resolution channels can help in reducing the contaminations from other sources and also improve the systematic errors in the measurement. }
\item{Measurement techniques should be devised such that the coupling of the FSD spectrum of the signal with the eigenvectors corresponding to the largest eigenvalues of the noise covariance matrix is minimized. This can improve the measurability of the spectral distortion signal and reduce the contamination from systematic error and instrumental noise.}
\item{The relative difference in the gain factor $G$ between frequency channels should satisfy the condition given in Eq. \eqref{extract-3a}.}
\item{A stable gain factor $G$ for the complete frequency range is required with a very small relative calibration error of $\delta G/G$ between different frequencies. The requirement for a controlled gain error is provided in Sec. \ref{sys-error}}.
\item{The systematic errors due to off-set temperature of the detectors must be controlled below the desired signal $S^\mu_\nu$ and $S^y_\nu$ at every frequency channel.}
\end{enumerate}

\section{Conclusions}
\label{conclusions}
The rich domain of cosmological information embedded in the spectral distortion of the CMB spectrum is going to be unveiled by the next generation of CMB missions. The CMB absolute intensity is usually compared with an internal blackbody calibrator to search for any deviations from blackbody. In the presence of an internal blackbody calibrator, the observed intensity of the sky is compared at every frequency with the intensity from the internal blackbody calibrator and any departure of the observed sky intensity from the blackbody can be modeled with the known spectrum of spectral distortions. As a result, a successful measurement of spectral distortion signals with a high SNR requires the internal blackbody calibrator to be extremely stable at a fixed temperature and also should obey a perfect blackbody spectrum over the complete frequency range of a mission (typically $1-1000$ GHz). The departure of the internal calibrator from blackbody can induce a systematic error and can also be misunderstood with the spectral distortion signal. 

We propose an alternative strategy called the Frequency Space Differential (FSD) to measure spectral distortions in  CMB. This technique measures the difference in the observed brightness temperature at different frequencies and models the observed difference with the theoretically predictable FSD kernel for different components in Eq. \eqref{source-2a}. The FSD spectrum for expected sources of spectral distortions like $\mu$ \& $y$ are different and not degenerate, which makes it easily distinguishable and extractable. The $\mu$ spectrum is stronger at low frequencies and decreases rapidly at higher frequencies, whereas $y$ distortion FSD spectrum is dominant at high frequency range as depicted in Fig.~\ref{cosmo-spectrum}. 

Our proposed method uses the CMB itself between the neighboring channels as a calibrator to measure the deviations from blackbody. This method  does not directly measure the absolute blackbody spectrum, but only measures the frequency space derivative of a blackbody signal. In the presence of spectral distortions, the FSD signal exhibits a combination of effects from blackbody along with other sources and can be fitted uniquely for a known FSD spectrum. Successful implementation of the FSD method needs several instrumental controls in order  to reduce contaminations by systematic errors and instrumental noise, which we listed in Sec.~\ref{requirement}. Measurement of the spectral distortion signal without an internal absolute blackbody calibrator can be possible in implementing  this formalism via suitable instrumental engineering for  future missions. 

The main insight of this paper is to explore signatures of spectral distortion and measuring any deviations from blackbody through the FSD spectrum. This process enables one  to measure the spectral distortion signal in  the same spirit as WMAP measured  the CMB anisotropies through a differential measurement without an internal reference. The main advantage of our method is that it does not require an internal blackbody calibrator to measure the signal. Secondly, this approach opens up an alternative way of measuring the spectral distortion signal which can be  useful for comparing results from other missions which use an internal blackbody calibration method. Next-generation CMB missions with  upgraded detector technologies can implement this method to measure spectral distortions without using an absolute calibrator.  Estimation of the noise properties and experimental requirements in order  to implement this method for a future CMB mission like LiteBIRD \citep{Matsumura:2016sri}   will be addressed in a follow-up paper. 
 
 \textbf{Acknowledgements}
This work has been done within the Labex ILP (reference ANR-10-LABX-63) part of the Idex SUPER, and received financial state aid managed by the Agence Nationale de la Recherche, as part of the programme Investissements d'avenir under the reference ANR-11-IDEX-0004-02. The work of JS has been supported in part by ERC Project No. 267117 (DARK) hosted by the Pierre and Marie
Curie University-Paris VI, Sorbonne Universities. The work of SM and BDW is supported by the Simons Foundation. The authors acknowledge valuable comments from Masashi Hazumi on this draft.

\bibliographystyle{mnras}
\bibliography{spectral-distortion-derivatives-Final-mnras_v2}
\bsp	
\label{lastpage}
\end{document}